\pdfoutput=1 
\documentclass[11pt, letterpaper, bookman]{article}
\usepackage[margin=1in]{geometry} 
\usepackage{graphicx}
\usepackage{tikzpeople}

\usepackage{booktabs} 
\usepackage[font=small,labelfont=bf]{caption} 
\usepackage{wrapfig} 
\usepackage{tabularx}
\usepackage{bbm}
\usepackage{thm-restate}
\usepackage[square,numbers,sort&compress]{natbib}
\usepackage{url}
\usepackage{authblk}
\usepackage[capitalise,noabbrev]{cleveref}
\usepackage{float}
\usepackage{enumitem,paralist}
\usepackage[T1]{fontenc}

\AtBeginDocument{%
	\providecommand\BibTeX{{%
			\normalfont B\kern-0.5em{\scshape i\kern-0.25em b}\kern-0.8em\TeX}}}

\begin{document}

\title{Fairness in Agreement With European Values: An Interdisciplinary Perspective on AI Regulation}

\author[1]{Alejandra Bringas Colmenarejo}
\author[2]{Luca Nannini}
\author[3]{Alisa Rieger}
\author[4]{Kristen M. Scott}
\author[5]{Xuan Zhao}
\author[6]{Gourab K. Patro}
\author[7]{Gjergji Kasneci}
\author[8]{Katharina Kinder-Kurlanda}
\affil[1]{University of Southampton, United Kingdom}
\affil[2]{Minsait - Indra Sistemas and CiTIUS, Universidade de Santiago de Compostela, Spain}
\affil[3]{Delft University of Technology, Netherlands}
\affil[4]{KU Leuven, Belgium}
\affil[5]{SCHUFA Holding AG and University of Tuebingen, Germany}
\affil[6]{IIT Kharagpur, India and L3S Research Center, Germany}
\affil[7]{SCHUFA Holding AG and University of Tuebingen, Germany}
\affil[8]{Digital Age Research Center, University of Klagenfurt, Austria}

\maketitle
\begin{abstract}
With increasing digitalization, Artificial Intelligence (AI) is becoming ubiquitous. 
AI-based systems to identify, optimize, automate, and scale solutions to complex economic and societal problems are being proposed and implemented. This has motivated regulation efforts, including the Proposal of an EU AI Act.
This interdisciplinary position paper considers various concerns surrounding fairness and discrimination in AI, and discusses how AI regulations address them, focusing on (but not limited to) the Proposal. 
We first look at AI and fairness through the lenses of law, (AI) industry, sociotechnology, and (moral) philosophy, and present various perspectives.
Then, we map these perspectives along three axes of interests:~\textit{(i) Standardization vs. Localization},~\textit{(ii) Utilitarianism vs. Egalitarianism}, and~\textit{(iii) Consequential vs. Deontological ethics} which leads us to identify a pattern of common arguments and tensions between these axes. 
Positioning the discussion within the axes of interest and with a focus on reconciling the key tensions, we identify and propose the roles AI Regulation should take to make the endeavor of the AI Act a success in terms of AI fairness concerns.
\end{abstract}
\section{Introduction}\label{sec:intro}
AI applications have grown at an unprecedented rate in recent years and have become ubiquitous in our society. 
While often deployed with the intention to increase efficiency and fairness of decision-making, AI has also sparked many debates on (un)fairness~\cite{sartori2022sociotechnical}. These debates surround, amongst others, unfair treatment of individuals and groups due to the reproduction of systemic, institutional, and societal biases in AI decisions~\cite{ caliskan2017semantics}; the opacity of AI decisions~\cite{ananny2018}; diverse jeopardies to democracy and societal well-being~\cite{muller2020impact}; risks to consumer privacy~\cite{manheim2019artificial}; and market inequalities that are observed in the aggregation of unprecedented levels of power of big companies that develop AI systems~\textit{(Big Tech)} while small and new companies are struggling to enter the market~\cite{santesteban2020big}.
In many fields of AI application, such as policing, justice, and recruitment, bias and unfairness as described above should not only be mitigated to increase fairness but in fact, to avert violating protected human rights. 

The above mentioned undesired effects and consequences of AI application and development propelled the European Union for new regulations, ex-ante reviews, and ex-post monitoring on AI systems. The European Union intends to assert the AI Regulation through the protection of human dignity and fundamental rights with the Proposal of the~\textit{Artificial Intelligence Act}~\cite{aiact}, convinced that human beings should remain at the center of technological development.
However, to make this endeavor of the AI Act a success, to some extent divergent interdisciplinary views and perspectives on bias, fairness, and regulation, have to be taken into consideration.

We elaborate on ~\textit{legal},~\textit{industrial},~\textit{sociotechnical}, and~\textit{philosophical} perspectives in light of identified axes of tension in the debate on AI fairness and regulation:~\textit{Standardization vs. Localization},~\textit{Utilitarianism vs. Egalitarianism}, and~\textit{Consequential vs. Deontological}.
Further, we discuss discrepancies between how these perspectives are addressed in the current Proposal of the Artificial Intelligence Act and make recommendations how they could be addressed for better reconciliation with all three perspectives and the legal requirements. 
In sum, we make the following contributions to the ongoing discourse on AI fairness and regulation: 
\begin{inparaenum}[i.]
    \item \textbf{Interdisciplinary perspectives:} Comprehensive interdisciplinary (technical, legal, industrial, sociotechnical, philosophical) discussion of bias, fairness, and regulation (\cref{sec:techframe,sec:legal,sec:industry,sec:sociotechnical,sec:philosophical}),
    \item \textbf{Mapping tensions of debate:} mapping the different perspectives on fairness in AI applications and regulation on to three axes that reveal tensions in the debate:~\textit{Standardization vs. Localization},~\textit{Utilitarianism vs. Egalitarianism}, and~\textit{Consequential vs. Deontological ethics} (\cref{sec:insufficiency}),
    \item \textbf{Path forward:} Recommendations towards consensus for a successful AI Act that reconciles divergent perspectives (\cref{sec:path_forward}).
\end{inparaenum}
\section{Technical Frameworks for Bias and Fairness in AI}
\label{sec:techframe}
In this section we present examples of fairness controversies for selected AI application domains with high-stake consequences. Subsequently, we discuss several AI fairness notions and present research on guidance to choose between these notions and between measures to mitigate bias in AI systems. 
\subsection{Examples of Bias and Unfairness in AI Applications}\label{sec:applications}
Automated decision-making systems were suggested to be capable of increased fairness due to avoidance of human bias interference \cite{kleinberg2017inherent}. However, many cases have come to light in which automatic decision-making was found to raise critical issues regarding fairness, and reproduces systemic, institutional, and societal biases. Such biases can result in discrimination, unfairness, and issues of privacy, thus, violating protected human rights (see~\cref{sec:legal}).
This is especially harmful when automated decision making has high-stake implications for individuals and society. In the following, we present salient examples.
 
In \textbf{Policing and Justice}, AI systems are applied across Europe to inform and assist day-to-day police work by profiling people, attempting to predict likely future behavior or locations of future crimes, and assessing the alleged risk of criminal involvement of individuals (e.g., \textit{Top 600 criminals list} and~\textit{CAS} (Netherlands),~\textit{Delia} (Italy),~\textit{SKALA} (Germany). 
Outcomes of these predictions and assessments are used to justify surveillance, searches, or questioning of alleged~\textit{high risk} individuals. However they have been suspected to reinforce existing patterns of offending and enforcement~\cite{adensamer2021part, sandhu2021uberization}.
In the judicial arena, automated decision-making is currently being applied in various courts around the world to support certain tasks, such as risk assessment of recidivism, as well as decisions concerning bail amounts, probation periods, and sentencing~\cite{re2019developing, zavrvsnik2020criminal}. Across Europe, such systems are not yet used widely, however, they have been introduced or tested in some countries, e.g., in Spain~\textit{(RisCanvi)} or the UK~\textit{(HART)}.
\citet{zavrvsnik2020criminal} highlights potentially violated rights due to opaque, automated decision-making in the justice system, e.g., the right to a fair trial, the principle of non-discrimination and equality, and the right for explanation. 

 AI systems are further being applied in the domain of {\bf Education and Employment}, to support candidate selection for higher education admissions and recruitment, e.g., with CV screening, targeted job advertisement, candidate sourcing, and video
 screening~\cite{albert_2019}. The risk of bias has been demonstrated at each of these stages in the recruitment process \cite{bogen2018HelpWantedExamination,lambrecht2019algorithmic}.

In {\bf Finance and Banking}, AI algorithms constitute the basis of numerous different applications, such as market forecasting for trading, or risk management for credit scoring, loan allocations, and mortgage rates~\cite{cao2022ai}.
Various cases have come to light in which decisions of such applications were found to be unfair and biased towards minority borrowers, i.e., with higher mortgage and loan rejection rates for Hispanic and Black borrowers in the US~\cite{bartlett2019consumer, finocchiaro2021bridging}, or lower credit limits for women than for men with equal credit relevant characteristics~\cite{telford2019apple, gupta2019algorithms}.
     
  For {\bf Online Platforms}, AI based recommender systems are applied to support users to navigate the web by filtering information and suggest items (videos, social media content, products, music,..) predicted to be relevant for the user. 
  Recommender systems were found to amplify different kinds of bias, such as~\textit{representation} bias with an over-representation of male, white, and young users~\cite{ribeiro2018media}, and~\textit{exposure} bias where the top 20\% of businesses get 80\% of the exposure~\cite{patro2020fairrec}, and marketplaces preferentially recommend their own products~\cite{dash2021umpire}. 
  This amplifies substantial power imbalances between market-dominating platform incumbents~\textit{(Big Tech)} and smaller platforms who do not have access to equal vast amounts of high-quality consumer data that is vital to enter the market~\cite{santesteban2020big}.
  The resulting immense power concentration in the private hands of very few companies that develop most AI applications and prioritize profit over benevolence for society poses an additional threat to democracy and society~\cite{simons2020utilities, epstein2019google}. 
  Further, recommender systems and search result rankings that often optimize to capture attention, determine a large extent of the information to which people are exposed. 
  This can result in distorted exposure to information and viewpoints, as well as exposure to dis- and misinformation, raising issues of fairness and posing a threat to democracies that are reliant on well-informed citizens who can engage in healthy political and social discourse~\cite{hillsDarkSideInformation2019, muller2020impact}. 
  AI systems could threaten democracy and society further by undermining the process of elections through targeted advertisements. Such~\textit{microtargeting} provides tools for interference by malicious political actors~\cite{dobber2019regulation, mullainathan2018algorithmic}.

\subsection{Mitigating Bias and Ensuring Fairness}
Most fairness definitions consider either group or individual fairness. \textit{Group fairness} is focused on requiring that people who belong to protected groups receive on average the same treatment/outcome as the overall population, expressed as the equality of a selected statistical measure across groups~\cite{verma}, such as \textit{statistical parity, demographic parity, equal opportunity} and~\textit{equality of odds}. \textit{Individual fairness} focuses on ensuring that any two individuals who are similar except for the protected features receive equal or similar treatment/outcomes~\cite{dwork2012fairness}.
While ideally, multiple fairness notions would be met to reach a~\textit{complete} fairness status, this is impossible due to mathematical incompatibilities between them~\cite{mitchell}.
Criteria to systematize the procedure of selecting between fairness notions when making a specific decision have been proposed: Amongst others, the existence of a ground-truth, base-rates between sub-groups, the cost of misclassification, or the existence of government regulations to meet may be considered~\cite{makhlouf2021applicability}. 

Formalization of fairness definitions in a specific context is nuanced and it is important that AI practitioners receive some guidance when designing a fair AI system. Some recent research proposes the~\textit{Fairness Compass}, a schema in form of a decision tree which simplifies the selection process by settling for the desired ethical principles in a formalised way~\cite{ruf}. A~\textit{standardized roadmap} could potentially make the identification of an appropriate fairness definition a more straightforward procedure, and help document the decision process toward fairness. Audit, monitoring and explanation might then be more accessible and less expensive. Nevertheless, there should also be space for stakeholders with deeper understanding of the specific context to contribute refinement and interpretations of any such roadmap.

The fairness notions mentioned above deal with the outcome of automated decision-making.
Counterfactual fairness~\cite{kusner2017counterfactual} and causal fairness~\cite{vonkugelgen}, however, have a procedural implication which might be more suitable for the cases where a counterfactual or causal connection needs to be established between features. Most of the existing fairness notions are formalized in a static scenario. If we want to better understand how bias is encoded in historical data or evaluate the consequences of certain fairness intervention, dynamic fairness notions~\cite{zotero-686} might offer a better solution.

Technical methods to mitigate bias in algorithms fall under three categories: (1) \textit{Pre-processing}. Pre-processing techniques try to transform/re-balance the data so that the underlying discrimination is mitigated;
(2) \textit{In-processing}. The construction of objective function usually has Utilitarian motivation behind, e.g. trying to maximize the utility of whole population. In-processing methods for bias mitigation can be used either by incorporating changes into the objective function or imposing a fairness constraint;
(3) \textit{Post-processing}. Post-processing methods reassign the labels initially predicted by the black-box model to a fairer state. ~\cite{mehrabi2021}.
 
The existing technical solutions toward fairness focus on more consequential approaches: the outcome/decision is evaluated by a specific fairness notion and then measures are taken to correct the unfair outcome/decision.
Concerns have been voiced that fairness cannot be simply achieved through mathematical formulation approaches as the \textit{formalism trap}~\cite{zotero-637} and the seeming success of these technical solutions in the end will hinder pursuits of actual fairness with the cooperation of social practices~\cite{harcourtd}. 
\section{A Legal Perspective on Bias and Fairness in AI}\label{sec:legal}
To follow one central goal of the EU---the promotion of peace and well-being for its members---EU law aims at ensuring that EU member-states and individuals are treated and treat each other equally and fairly. The blindfolded \textit{Justicia} further emphasizes the importance of laws that promote fairness, but also fairness within the enforcement of all laws.
Decision-making based on machine-learning could be a promising support for that, to mitigate the unconscious or deliberate biases that we as humans have. 
However, being trained on (biased) data from previous decisions, the promise of unbiased assessments could not be fulfilled so far~\cite{ hubner2021two, angwin_larson_mattu_kirchner_2022}.

In this section, we will take a structured look at the legal perspective on bias and fairness in AI. We will start with an overview of EU legislative framework on non-discrimination and the approach to fairness followed by the EU Data Protection Law. Then we will conclude by addressing the technical requirements to deal with bias that would be introduced with the AI Regulation Proposal. 
\subsection{Non-Discrimination Law}
The general principle of non-discrimination in EU law protects people from discrimination and unfair treatment. European anti-discrimination law is designed to prevent discrimination against particular groups of people that share one or more characteristics---called protected attributes---and from which the group acquires the category of a protected group. Concretely, protected attributes under the Charter of Fundamental Rights of the European Union include sex, race or ethnic origin, colour, ethnic or social origin, genetic features, religion or other belief, disability, age, sexual orientation, political or any other opinion, language, membership to a national minority, property, social origin, and birth (Art. 21.(1)) \cite{charter2007}. Additionally, the Charter prohibits discrimination on the grounds of nationality, compels the European Union to ensure the equality of everyone under the European law, demands the respect of cultural, religious, and linguistic diversity, and seeks equality of men and women in all areas. Several other European anti-discrimination directives have further covered the legal protection offered to these protected attributes. Specifically, under the European Legislation men and women must receive equal treatment in the labour market and regarding the access and supply of good as services\cite{directive2004,directive2006}. Likewise, equal treatment must be guaranteed between  persons irrespective of their racial or ethnic origin \cite{directice2000a}, as well as equity shall be respected in employment and occupation in regards to the grounds of disability, religion or belief, age and sexual orientation \cite{directive2000b}. Member States expanded the protection towards discrimination through specific national laws and provisions.

Furthermore, the European legislation presents two tools to address discrimination, \textit{direct} and \textit{indirect} discrimination. Direct discrimination is defined as a \textit{situation in which one person is treated less favourable on the grounds of a prohibited criterion than another is, has been or would be treated in a comparable situation} \cite{directice2000a}. Thus, it is straightforwardly related to the possession of a protected attribute that distinguishes the person from other individuals, regardless of the intention behind the disparate treatment or the mere existence of less favourable treatment. 
In the context of data-driven systems, direct discrimination will cover those cases where the model is not neutral towards a protected attribute and offers a less favourable output to individuals on the basis of protected groups, whether they truly fit into that group or are associated with the protected attribute. Since consciously inputting discrimination into the model will affect its accuracy, these cases are not of great concern \cite{xenidis2019eu}. 
 
By contrast, indirect discrimination will more likely capture many situations of algorithmic discrimination because it affects situations \textit{where an apparently neutral provision, criterion or practice would put members of a protected category at a particular disadvantage compared with other persons unless that provision, criterion or practice is objectively justified by a legitimate aim and the means of achieving that aim are appropriate and necessary} \cite{directice2000a}. Nevertheless, the prohibition of indirect discrimination does not encompass a set of clear and easily applicable rules, it can rather be considered closer to a standard than to a rule \cite{zuiderveen2018discrimination}. \textit{The concept of indirect discrimination results in rather open-ended standards, which are often difficult to apply in practice. It needs to be proven that a seemingly neutral rule, practice or decision disproportionately affects a protected group} \cite{zuiderveen2018discrimination}. Due to this, indirect discrimination concerns neutral models, which in principle are blinded to sensitive attributes or do not operate on the basis of those protective attributes. Thus, direct discrimination focuses on individual cases of discrimination, while indirect discrimination deals with rules and patterns of discrimination and can reveal underlying social inequalities.
\subsection{Data Protection Law}
The European Union General Data Protection Regulation (GDPR)~\cite{GDPR2016} refers to automated individual decision-making and seeks, amongst other objectives, to prevent algorithmic discrimination. Generally, the GDPR states the objective to protect all the fundamental rights recognised under EU law, which the processing of personal data may challenge. According to the GDPR, the core principles that shall lead the processing of personal data are lawfulness, fairness, and transparency. Concretely, the principle of fairness entails the processing of personal information that is not in any way \textit{unduly detrimental, unexpected, or misleading to the individuals concerned} (\cite{ICOreport}). Indeed, the principle of fairness seeks to protect the individual's fundamental rights and freedoms, and so, their non-infringement by such processing. Likewise, the \textit{principle of data accuracy} requires the control of the quality of data for its processing, although it does not address the possible wrongful or disproportionate selection of data and therefore the effect and consequences resulted from such selection~\cite{Ntoutsi2020}.
To ensure fair processing, the GDPR requests the use of \textit{appropriate mathematical and statistical procedures for profiling that take into account the risks involved for the interest and rights of data subjects and prevent discriminatory effects on natural persons} (Recital 71~\cite{GDPR2016}). Furthermore, the GDPR highlights the potential \textit{risks to the rights and freedom of natural persons, which could lead to physical, material or non-material damage, in particular when processing results in discrimination} (Recital 75~\cite{GDPR2016}). Despite these provisions, ensuring fairness is still quite a subjective matter as it requires that the data processing shall not exceed reasonable expectations nor provoke unjustified adverse effects on the individuals. However, what can be considered reasonable expectations and justifiable effects is an open question, leaving the notion of \textit{fair processing} undefined. 

However, the European anti-discrimination law evidently embedded notions of substantive discrimination and therefore, unjustified algorithmic discrimination, as referred to in Article 5 and Recital 71, implies unfair processing~\cite{hacker_2018}. From the legal perspective, discrimination collides with equality, infringing the principle of fairness; whereas from a technical perspective, algorithmic discrimination straightforwardly entails unfair processing (see~\cref{sec:techframe}). 
\subsection{EU Artificial Intelligence Regulation Proposal}
With the EU Artificial Intelligence Act the European Union aims at laying down harmonized rules on artificial intelligence with four specific objectives~\cite{aiact}: \textit{
\begin{inparaenum}[1)]
    \item ensure that AI systems placed on the Union market are safe and respect existing law on fundamental rights and Union values;
    \item ensure legal certainty to facilitate investment and innovation in AI;
    \item enhance governance and effective enforcement of existing law and safety requirements applicable to AI systems;
    \item facilitate the development of a single market for lawful, safe and trustworthy AI applications preventing market fragmentation. 
\end{inparaenum}}

In essence, the Proposal seeks to balance legal certainty and the development of AI systems while ensuring an approach that respects European values, principles and laws. The specific purpose of the Proposal is to establish a classification for trustworthy AI systems based on a risk-based approach, to introduce new legal obligations and requirements on public authorities and businesses for the development and application of AI systems, to prohibit harmful AI-enabled practices, and to set new monitoring and enforcement regimes. Essentially, the Proposal will set a legal framework applicable for developers and end-users of AI systems which \textit{specific characteristics---opacity, complexity, dependency on data, autonomous behaviours---can adversely affect a number of fundamental rights enshrined in the EU Charter of Fundamental Rights} \cite{aiact}.

The Proposal delimits a set of prohibited AI practices considered harmful because they contravene EU values and violate fundamental rights. Second, the Proposal outlines specific obligations to avoid the appearance of bias in two types of high-risk AI systems; (1) those which are intended to be used as a safety component of a product or is itself a product, and this product is subject to an existing third-party conformity assessment, and (2) those which are involved in decision-making processes in the following areas; (i) biometric identification and categorization of natural persons, (ii) management and operation of critical infrastructure, (iii) education and vocational training, (iv) employment and workers management as well as access to self-employment, (v) law enforcement, (vi) migration, asylum, and border control management, and (vii) administration of justice and democratic processes (see~\cref{sec:applications}).

According to the Proposal, AI systems can only be placed into the EU market if they comply with the certain minimum requirements specified in the legislation, requirements that become stricter as the risk associated with the system increases (i.e., minimal risk, low risk, high risk, and unacceptable risk). Consequently, providers will need to carry out ex-ante conformity assessments and implement quality and risk management systems and post-market monitoring to ensure compliance with the new regulation and minimise the risk for users and affected persons. However, the Proposal pays little attention to identifying the causes and proposing recommendations to tackle the potential discriminatory harms of AI systems. Specifically, the Proposal mainly focuses on biases in data sets, forgetting other types such as those that may arise from the choice of algorithms, and the optimization or evaluation of metrics. Additionally, the Proposal may pose unreasonable trust in human operators---i.e., human in the loop---to identify and recognise cases of bias and discrimination in AI systems.

The Proposal does not provide detailed guidance on dealing with unavoidable trade-offs for the different stakeholders when debiasing and monitoring bias in the data set. Nevertheless, some insights can be found in the Proposal regarding the expected requirements to debias high-risk AI systems. Firstly, there will be an obligation to establish appropriate data governance and management practices concerning the training, validation, and testing of data sets, in particular, to examine possible biases, ensure the relevance, representativeness, absence of errors and completeness of the data sets, and their consideration with the characteristics or elements that are particular to the specific geographical, behavioural or functional setting within which the high-risk AI system is intended to be used \cite{aiact}. Secondly, a novel exception to the Data Protection Regulation will allow \textit{to the extent that it is strictly necessary for the purposes of ensuring bias monitoring, detection and correction in relation to the high-risk AI systems} \cite{aiact} the processing of special categories of data. Finally, the Proposal asks for developing methods that will ensure the detection of biased outputs and the consequent introduction of appropriate mitigation measures as it recognises the potential of AI systems to develop biased outputs due to outputs used as an input for future operations, i.e., \textit{feedback loops}.

Interestingly, the Proposal also details the role of standards and specifications in the AI landscape \cite{aiact}. On the one hand, the Proposal addresses the use of \textit{harmonised standards} to presume the conformity of AI systems with the regulation's requirements. On the other hand, the Proposal entitles the Commission with the duty to adopt common specifications and technical solutions when the harmonised standards are insufficient or there is a need to address specific or fundamental rights concerns. In other words, \textit{conformance with technical standards and common specifications should give providers of high-risk AI a level of confidence that they are compliant with the mandatory requirements of the proposed EU AI Regulation as well as significantly cutting the cost of compliance for business} \cite{mcfadden2021harmonising}. Whereas neither the standards nor the specifications will be compulsory for providers of high-risk AI systems, their non-adoption shall entail a justification as to which and why other technical solutions were adopted. 
\section{An Industry Perspective on Bias and Fairness in AI} 
\label{sec:industry}
Substantial research on ML fairness, even for industry applications, has originated out of academic contexts. Academic research has first proposed most fairness principles and quantitative methods to mitigate biases and unbalanced data with general application domains \cite{mehrabi2021, ayling2021putting, lee2021landscape}.
Toolkits appeared ready to be integrated for the industry, even if often developed following non-contextual design rationales based upon the issues of algorithmic methods \cite{hoffmann_where_2019}. 
Until recently, the technical nature of academic contributions have often not addressed the practical issues that industry practitioners face when adopting and engaging with fairness tools. Practitioners have pointed out the lack of ethical tools' usability in real-world applications due to a series of critical factors preventing the straightforward adoption of fairness principles and methods \cite{mittelstadt_principles_2019}. Following \citet{morley_ethics_2021}, such non-effectiveness in real-world cases stems from how fairness compliance is operationalized inside companies. If not developed with the sociotechnical features and constraints of AI product deployment in mind, these methods could easily lead to failures \cite{hoffmann_where_2019} including for example fairness definitions misinterpretation \cite{krafft_defining_2019}, obfuscation of practitioners' accountability \cite{orr2020attributions}, and gaming fairness measures as a method of ethics-washing \cite{morley_ethics_2021}. 
To avoid shortcomings, researchers are now focusing on how to operationalize fairness frameworks based on the needs of industry practitioners.
\citet{veale_fairness_2018} conducted interviews with decision makers in high-stakes public-sector contexts. Practitioners were found to be lacking incentives and practices for algorithmic accountability due to resource constraints and dependency on prior infrastructure. 
\citet{holstein_improving_2019} enlarged the pool of industry practitioners with a systematic investigation of ML product development. Amid the area of intervention were identified issues of data quality provenance and reporting, as well as the need for domain-specific educational resources and compliance protocols, intended specifically as internal auditing processes and tools for fairness-focused debugging. \citet{rakova_where_2021} reported that practitioners often felt a hostile organizational environment where they were hindered or uncompensated when trying to implement fairness practices independently. Disincentive stems from the lack of educational programs, rewards, accountability allocation, and communicative protocols over fairness issues, especially when different parts of an AI development are distributed across different teams. This resulted in practitioners often feeling disoriented, unprepared, or even overwhelmed by fairness tools and checklists \cite{cramer2019translation, holstein_improving_2019}. 
It was also observed that practitioners recommend establishing internal and external investigation committees to create an inclusive and preventive environment and to provide resources such as protocols or educational teams \cite{madaio2020co, rakova_where_2021}. Other research examples, once informed on practitioners' needs, focused on designing different AI fairness solutions: checklists to be aligned with teams' workflows and organizational ad-hoc processes, fairness frameworks or internal algorithmic auditing protocols designed for industrial applications \cite{madaio2020co, raji_closing_2020}. Recently, \citet{richardson_framework_2021} proposed a complete industry framework of stakeholders and fairness recommendations while specifying operationalization pitfalls. 
\citet{ibanez_operationalising_2021} distinguished two main perspectives on operationalizing fairness practices in organizations: a bottom-up, reactive approach, where prior organizational processes restrain best practices, or top-down, where a proactive approach is set in place according to the translation of principles and methods as actionable, iterative steps designed with stakeholders' needs and concerns in mind.
Interestingly, the literature agrees that fairness interventions should not be standardized and reactive to prior single instances of organizational infrastructure issues, but proactive, based on a thorough understanding of different stakeholders' needs, and accounting for domain-specific and contextual factors.

In regards to the Proposal, it is not yet clear how fairness practices will be effectively operationalized given the mechanisms envisioned in Articles 43 and 61 from the Proposal, respectively for conformance checking and post-market monitoring of high-risk systems. For those systems, providers will be demanded to draft and verify their conformance through a \textit{quality management system}, \textit{technical documentation}, and \textit{post-market monitoring} under the lens of a national body. This body will be guided by a national supervisory authority in coordination with the EDPB (European AI Board from the EU commission). Yet, some detractors, in line with some concerns over organizations' ethics washing, advanced skeptical doubts on the procedural efficacy of these auditing mechanisms  \cite{propp_machines, manners_normative_2002}. Doubts were related to the undisclosed nature of conformity declarations as well as the nature of contributions of data criteria input to the \textit{EU database for stand-alone high-risk AI systems} in Article 60, withheld from the scrutiny of those affected by such systems and available only upon regulatory bodies' request. This loose gravity towards the public interest might not permit to enforce EU citizen fundamental rights to decide whether a system should be listed as high-risk.  
In light of the concerns for more structural fairness practices, the evolution of an overly rigid and costly compliance environment could critically undermine these needs. An official impact assessment has been proposed \cite{renda2021study} to quantify these costs. 
\citet{mueller2021much} advanced an analysis of the economic costs that could arise for EU small and medium enterprises and corporations. 
In the forecast, effects will push away venture capital investors, drain European talents and tighten stronger external dependencies leading to a highly unfavorable European environment, with the risk of being excluded from the global AI market. Academics and policy analysts have advanced a debate on the validity of those claims, picturing less-burdening assessments over quality management systems, thus calling the report factitious \cite{RePEc:osf:socarx:8nzb4, noauthor_clarifying_2021}. 
Future predictions will need to account both for amendments to the terminology and procedures. Foremost, central analysis focus should be given to the ecosystem of digital technology regulations that the EU has on its agenda \cite{European_Commission_2022}. 
These digital Proposals constitute the European intention of enforcing its legislative sovereignty and set standards for the international market. Leveraging the \textit{Brussels Effect} \cite{bradford2020brussels, greenleaf_brussels_2021} and the current rise of AI ethics attention across a wide range of institutional and academic stakeholders \cite{schiff_global_2022, gupta2022state}, it is reasonable to predict that in the near future current investments in integrating fairness governance practices could be streamlined into more mature and efficient regulatory frameworks with lower procedural costs while mitigating reputational risks \cite{rakova_where_2021}. 
\section{A Sociotechnical Perspective on Bias and Fairness in AI}
\label{sec:sociotechnical}

Regarding AI fairness and discrimination, many have pointed out that AI is not merely a tool, it is a sociotechnical endeavour, meaning that the development, use of (and harm from) AI technologies can not be separated from their specific social contexts \cite{elishSituatingMethodsMagic2018,penagangadharan2019DecenteringTechnologyDiscourse}. When attempting to prevent harm from technologies we must look closely at a new technology's actual capacities and functions within these contexts. An over-emphasis of the role of specific technological features of AI in either causing, or preventing, discrimination, for example, can obscure other forms of discrimination that are occurring, as well as lead to an unproductive and ultimately distracting focus on~\textit{fixing} or regulating those specific features \cite{penagangadharan2019DecenteringTechnologyDiscourse, grgic2018beyond}.

\citet{vealeDemystifyingDraftEU2021} make a similar argument in regards to the Proposal. They cite the examples of the prohibition against releasing AI systems that use subliminal or subconscious techniques to distort a person's behaviour and argue that this focus on evocative,~\textit{ripped from the headlines} potential harms does little to mitigate actual harms and adds little to existing legislation \cite{vealeDemystifyingDraftEU2021}. Issues include, for instance, that prohibition only covers manipulative systems that cause individual harm but not a collective harm or a \textit{harm that arises from dynamics of the user-base entwined with an AI system}~\cite{vealeDemystifyingDraftEU2021} and that there must be intent to distort behaviour. 
\citet{pauldourish2011DiviningDigitalFuture} identified a similar phenomenon surrounding the discussion and implementation of ubiquitous computing technologies and contrast the \emph{myth} used to build visions of technologies and the \emph{messiness} of the practical implementation of technologies in reality. They further describe ubiquitous computing researchers as explaining away limitations and unexpected consequences of specific systems by referring to a proximate future where the given technology will be fully realized and highly useful, as soon as a few remaining kinks (such as unevenly distributed infrastructure, for example) are ironed out \cite{pauldourish2011DiviningDigitalFuture}. 

In the case of the \emph{messy} realities of AI, it is widely acknowledged that it is non-trivial to build \emph{error-free} models and good quality data within the context of societal factors and power structures at play \cite{miceli2020SubjectivityImpositionPower, elishSituatingMethodsMagic2018,chasalowRepresentativenessStatisticsPolitics2021}. To give a specific example, data workers who are frequently manually labeling, cleaning, and enriching the data used for training AI models, have a crucial role in the development of AI systems and their practices are subject to a myriad of non-objective influences \cite{miceli2020SubjectivityImpositionPower}. Similarly, the harms often identified with AI use online, such as hyper-personalization, invasion of privacy, and spread of hate speech can stem from issues beyond the technology, such as monopolies, data power imbalances, and un-checked corporate crime \cite{doctorow2021HowDestroySurveillance}. Some have argued that those aspects of online life are a requisite feature of an emerging economic system that has grown out from the existing capitalist economic system \cite{zuboff2019SurveillanceCapitalismChallenge}. 

Therefore, we must acknowledge the systemic sources of the discrimination when mitigating discriminatory harm of AI technologies and the discussion of the impact of such technologies should start at an earlier point. In particular, we must look at the specific setting of a given case. This includes considering what specific sociopolitical goals a given AI system is enforcing. For example, in Austria, a risk assessment algorithm created for use in the public employment system has been described as guided by a philosophy of neo-liberal austerity in the social sector which has been replacing the concept of the European welfare state \cite{allhutter_cech_fischer_grill_mager_2020}. We must also consider where the discussions are happening, who is involved in the discussions, and how the population is able to discuss and enforce whether an AI in a domain should be used at all. In regards to the Proposal, according to~\cite{vealeDemystifyingDraftEU2021}, there is evidence of industry influence in high level policy decision-making surrounding the current Proposal.

Another complication in regulating and mitigating harm from AI is the complexity of determining how, or if, it is possible to distinguish between AI decisions and human decisions. If we do not acknowledge these entanglements, there is a risk of bias being addressed with overly mechanistic approaches. In reference to the example of privacy ethics, \citet{nissenbaum2009PrivacyContextTechnology} has described how a focus on the very attempt to mitigate privacy concerns by ever more sophisticated anonymization methods can lead to overlooking other issues, such as algorithms that do not infringe on privacy, yet are still harmful. Similarly, a focus on attempting to operationalize a very specific concept of fairness, and to regulate specific methods for monitoring it, risks pulling awareness from other algorithmic harms, or even obfuscating underlying causes of harm \cite{balayanDebiasingRegulatingAI2021,penagangadharan2019DecenteringTechnologyDiscourse}.
In the case of the Austrian AMS, described above, the controversy of a proposed algorithm opened up a whole discussion about how a Public Employment System should be run overall. From the perspective of power aware analysis \cite{miceli2020SubjectivityImpositionPower} everyone affected needs to be involved in those decisions. 
\section{A Philosophical Perspective on Bias and Fairness in AI}\label{sec:philosophical}
We also look at developments in AI and algorithmic fairness through the lens of moral philosophy, specifically normative ethics \cite{kagan2018normative}, which essentially investigates the question of whether something is morally right or wrong.
There are two major schools of thought in normative ethics;
(i) {\it Deontological ethics} argues the existence and significance of inherent rightness of an action (examples include Kant's {\it categorical imperative} \cite{paton1971categorical}, and Rawls' {\it veil of ignorance} \cite{rawls2009theory});
(ii) {\it Consequentialism} judges the morality of an action based on the value it brings (examples include {\it welfarism} \cite{keller2009welfarism}, hedonism \cite{Moore2008-MOOH}).
While our deontological views inform the building blocks of morality in today's society (e.g., EU fundamental rights), consequential approaches enjoy scalability through the use of representative or proxy metrics in real-world usages (e.g., cost-benefit analysis \cite{layard1994cost} or per-capita income in economics, and overall accuracy in machine learning as discussed in~\cref{sec:techframe}).
Traditional AI research often follows a declarative approach where a mathematical objective is designed and optimized while caring less about the decision-making process and its correctness or representativeness \cite{carabantes2020black,burkart_huber_2021,finocchiaro2021bridging}.
Such an approach can be argued to be a consequentialist's approach to AI whereby only the optimization of final objective matters and the end justifies the procedure.
However, this approach has received a lot of critique within the AI domain, and a range of issues have been pointed out; for example concerning causality \cite{guo2020survey,castro2020causality}, fairness \cite{mehrabi2021,finocchiaro2021bridging}, explainability \cite{burkart_huber_2021}, including the comparability and robustness of explanations \cite{PawelczykBHRK21, PawelczykBK20}, and trustworthiness \cite{toreini2020relationship}.

Another angle from which AI developments can be looked at, is {\it Welfarism} \cite{keller2009welfarism} (a type of consequentialism), which suggests choosing the action that maximizes the welfare or well-being of the population.
In fact, it is widely used in some areas of economics, game theory, social-choice theory, and applications.
Welfarism is often studied in two major forms;
(i) {\it Utilitarianism} \cite{sen1979utilitarianism} emphasizes maximizing the welfare of the population; 
(ii) {\it Egalitarianism} argues for equality often leading to a form of Rawlsian justice \cite{rawls2009theory} which comes under deontological ethics, but its objective form in welfarism tries to maximize the welfare of the worst-off.
Utilitarianism is found to be heavily embedded in today's society.
For example, the optimization objectives (loss functions) in machine learning are often the aggregate errors over the set of data points or the individuals, i.e., utilitarian in nature. Utilitarian social welfare is quite prevalent in economics, computational social choice (allocation, voting, etc.)\footnote{Nash social welfare \cite{kaneko1979nash} is an exception.}.
Such utilitarian objectives tend to optimize for the overall utility while may be best-serving the majority and poorly serving the minority populations.
This is one of the reasons due to which the usual loss-minimizing objectives have been found to be unfair in many applications including criminal justice, banking, and gig-economy.
On the other hand, egalitarian welfarism in machine learning would likely try to equalize the errors of all or groups of individuals instead of minimizing the aggregate errors.
In fact algorithmic fairness notions like individual fairness \cite{dwork2012fairness}, equal opportunity and equality of odds \cite{hardt2016equality}, equal mistreatment \cite{zafar2017fairness} are either inspired by or promote egalitarian views in consequential modeling (error represents a consequence).
These notions have been found to reduce the effects of pre-existing biases in data and to improve the utilities of marginalized groups under algorithmic decision-making systems.

A few recent works have also explored non-consequential or deontological approaches to algorithmic fairness. 
These works can be grouped into two categories.
(1) Works on {\it procedural fairness} \cite{grgic2018human,green2019disparate} argue that it is essential for the chosen design and principles to be socially acceptable. Thus, these works focus on understanding how people assess fairness and ways to infer societal expectations about fairness principles thereby accounting for all voices in designing fair decision-making systems.
For example, \citet{grgic2018human} propose a framework for procedural fairness by evaluating the moral judgments of humans regarding the use of certain features and accordingly designing decision-making systems.
(2) Another set of works argue for causal and counterfactual fairness, i.e., addressing unfair causal effects of sensitive attributes in the decision-making process \cite{castro2020causality,kusner2017counterfactual}.
Instead of focusing on the outcome alone, these works have explored deontological aspects and propose to ensure fairness in the decision-making process.
\section{Mapping Perspectives}\label{sec:insufficiency}
\begin{figure}
    \centering
    \includegraphics[width=0.8\textwidth]{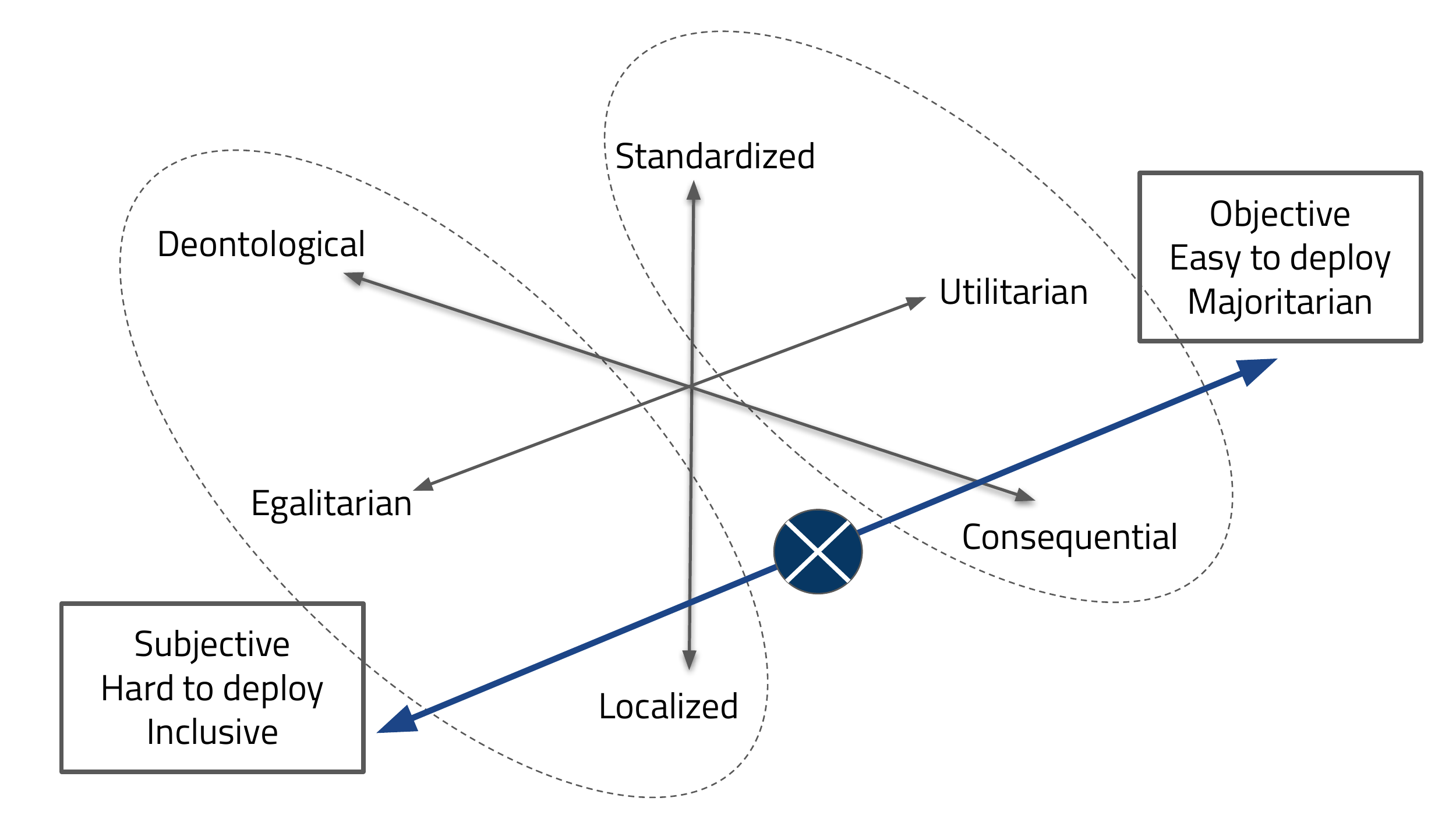}
    \caption{Three identified axes along which the debate about regulation of fairness in AI falls; Here they are aligned along high-level characterizations and common arguments made for, and against, each of the ends of the identified axes.}
    \label{fig:axis}
\end{figure}
We have identified three axes along which different perspectives in the debate about AI Regulation for preventing unfairness and discrimination fall. 
These axes may or may not be orthogonal, i.e., they may have relationships with each other. 
In the following sections, we define the axes and describe the debate surrounding regulating fairness in AI represented by each axis. These are not all of the axes of debate, rather these are salient tensions that we have identified. We find them helpful in conceptualizing and mapping the values and desiderata of the perspectives we are focusing on.
\subsection{Axis-1: Standardization vs. Localization}
\subsubsection{\bf The axis} This first axis of interest addresses the differences between standardization and localization. Standardization entails \textit{the process of making things of the same type all have the same basic features} (from Cambridge dictionary), specifically, through the creation of protocols to guide the design, development, and creation of such goods or services based on the consensus of all the relevant parties in the industry. Standardization is intended to ensure that all the goods and services produced respecting those protocols come with the same or equivalent quality, safety, interoperability and compatibility. For this reason, multiple parties need to be involved in developing such protocols and standards, namely, manufacturers, sellers, buyers, customers, trade associations, users or regulators (\url{https://www.iso.org/standards.html}). By contrast, localization describes \textit{the process of making a product or a service more suitable for a particular country, area, etc.} (from Cambridge dictionary). In essence, localization entails adapting the product or service to the characteristics of a given culture, region, or society. 
\subsubsection{\bf Pros and cons} 
In the context of AI, advocates for and members of industry frequently cite standardization as a method for preventing or mitigating discrimination \cite{Koene2018,Havens2018,discriminatoryAI}.
In this respect, high-risk AI systems will be presumed to comply with the requirements established in the AI Proposal if they are, as well, in conformity with the harmonised standards published by the Official Journal of the European Union as referred to in article 40 \cite{aiact}. Likewise, high-risk AI systems in conformity with the specifications referred to in Article 41 of the AI Proposal will be presumed in conformity with the regulation \cite{aiact}. In this sense, conformity with standards and specifications as proposed in the AI Regulation will allow the entry of high-risk AI systems in the European market while guaranteeing agreed levels of quality and safety that ensure the adherence to European principles and values (i.e., non-discrimination, fairness, and human dignity).  

A dilemma regarding standardization, however, appears when there is a disagreement regarding the standard of fairness that should be used to assess AI systems. As presented in~\cref{sec:applications} the straightforward example of incompatible fairness standards referred to the case of COMPAS and the different standards followed by ProPublica \cite{angwin_larson_mattu_kirchner_2022} and Northpoint \cite{Dieterich2016} for their fairness assessments, i.e., disparate mistreatment and calibration respectively \cite{zuiderveen2018discrimination}. Moreover, overly specific and strict standards and frameworks risk encoding a biased, restrictive, non-relevant to everyone, singular worldview, and may ultimately lead to uniformization from a top-down approach \cref{sec:industry}. 
In truth, standardarization as a method to enforce fairness can in some cases overlook the root-causes of bias, setting standards and notions of fairness that do not offer a real solution to the intrinsic discrimination or biases in certain situations or contexts \cref{sec:sociotechnical}. A---purely hypothetical---example of this problem would be the hard-coded requirements for gender parity in school admissions or hiring where there was a low representation of one of the genders, e.g., due to relocation for work reasons or armed conflicts. The solution would be to establish an acceptable ratio of males to females set at a level appropriate to the local context, rather than a strict gender parity requirement. 

In this regard, localizing AI systems entails the process of making them local in character by limiting the ethics regulation and specifics of enforcement to the desired area. Whereas the complete localization of AI systems will be in conflict with the embedded values of the AI Regulation (e.g., European Common Market and European Fundamental Rights), the localization of some of the decisions regarding their design, development, or deployment may allow a more tailored approach to address AI discrimination and biases in specific geographical, cultural, or sociotechnical contexts. The localization of some requirements and technical solutions may, as well, allow for the definition of ethical and legal guidelines that address the specific circumstances of a community, local area, or sector beyond the general standards and specifications. 
\subsection{Axis-2: Utilitarian vs. Egalitarian}
\subsubsection{\bf The axis} The second axis of interest addresses differences between utilitarian and egalitarian views. While a utilitarian philosophy is one of maximizing the overall welfare of the population, egalitarianism aims for equality amongst all those people.
\subsubsection{\bf Pros and cons} 
Utilitarianism has long been argued to be in conflict with the certain conceptualizations of fairness (see Chapter 14 of \citet{hooker_2014}).
In the context of AI, algorithms are often designed to optimize for certain mathematical objectives (which can be categorized as a declarative approach).
The objective functions in machine learning tasks usually measure a form of aggregate accuracy over a population, which fits the definition of a utilitarian measure. Optimizing solely for such a measure in AI applications risks optimizing the utility of the whole population while hurting minority groups in many  \cite{hardt2016equality,zafar2017fairness}. 
Utilitarian approaches are so ingrained in the computing research and development mindset that the early group fairness notions---which are supposed to mitigate the discriminatory effects of utilitarian objectives---such as demographic parity, had been reduced to utilitarian forms by constraining over the aggregate benefits or outcomes of groups of individuals \cite{zafar2017parity}.
The literature has now moved on to notions such as individual fairness, equal opportunity, and treatment parity which, even though outcome-based, are more egalitarian in nature.

Despite its obvious conflicts with fairness, and egalitarianism's close connection with fairness, utilitarian welfare is often cited a necessary factor in system and policy design. In fact, protecting the EU's economic interests is stated as a goal of the AI Act \cite{aiact}. 
Since utilitarianism captures a certain overall efficiency of a system (accuracy in machine learning, utilitarian welfare in economics), its goals often reflect business-oriented metrics of AI applications (i.e., click-through rate for recommendations in online marketplaces, or success-rate of ranked workers on gig-economy platforms). However, there might be a trade-off between maximizing efficiency and achieving other social objectives like equity or fairness in cases of inherent imbalance in the data or population \cite{berkconvex, bertsimas2012efficiency}.
\subsection{Axis-3: Consequential vs. Deontological}
\subsubsection{\bf The axis} This third axis of interest from the discussions in \cref{sec:legal,sec:industry,sec:sociotechnical,sec:philosophical} represents the differences between consequential and deontological ethics.
Deontological ethics argue for the existence of the inherent rightness of an action, while consequential ethics evaluate morality based on the consequences of an action. 

\subsubsection{\bf Pros and cons}
Technical measures for mitigating AI based discrimination tend to focus on fairness notions, whereby a fairness constraint is often added to the original objective. Fairness in this case is defined by statistical properties of the outcome/decision of the system (e.g., demographic parity). Fairness notions thus seek to reduce harm by adjusting or influencing the outcome to fit some statistical definition of fairness. While the motivation for doing this may be based on deontological principles of equality,
this approach belies a consequentialist definition of fairness, wherein one declares that fairness has been achieved through an equality in outcome, such as equal amount of good (accurate) and bad (inaccurate) outcomes for each group.

Deontological ethics is often given as an opposite to consequentialism. A deontological approach argues for the existence and significance of the inherent rightness of an action; in the context of AI based discrimination, this would suggest that the approach described above does not meet the criteria of acting morally, as the focus is on shifting the outcome. From a deontological perspective, an AI system is unlikely to be fair if the development of AI itself is not driven by essential guiding principles, such as fairness.

The Proposal's prohibition of certain uses is based on deontological principles of protecting fundamental individual rights. However, the risk based approach could be viewed as consequential, in that it only targets systems used in contexts perceived as being highly consequential. This means that many AI systems which might exhibit harmful representational or discriminatory biases, such as social media and online platforms are relieved of any requirements.

~\\{\bf Summary:} Based on the pattern of high-level characterizations and common arguments made for, and against, each end of the identified axes, we place them along a single axis, with one end containing localized, deontological, egalitarian approaches (LED) and the other end containing standardized, utilitarian, consequential approaches (SUC); we illustrate this mapping in Figure~\ref{fig:axis}. The LED end contains approaches that purport to acknowledge systemic and complex causes of discrimination and are often criticized as being overly subjective and hard to deploy. The approaches on the SUC end purport to be objective and easy to implement while often being critiqued as failing to recognize systemic causes or ensure inclusion of minority voices.
This mapping of the perceived benefits  and shortcomings of each approach allows us to identify a key tension in the debate on regulating fairness in AI. It is one that is based on differing understandings of the nature of bias and discrimination, along with differing priorities as to what constitutes practicality and implementability in efforts to increase fairness.
Following this, we suggest how the Proposal could better balance these values, as well as the differing perspectives of stakeholders, to achieve the stated goal of guaranteeing agreed levels of quality and safety in accordance with European principles and values (i.e., non-discrimination, fairness, and human dignity) without creating major hurdles for the European AI Industry.

\section{Key Agreement and A Path Forward}\label{sec:path_forward} 
\subsection{Key Agreement}
We see a specific agreement amongst the presented perspectives, regarding limitations of the current regulation. Ultimately each of the perspectives agree that regulation needs to be grounded in the reality of the context of the use of AI, and that this is not sufficiently achieved in the Proposal. A brief summary of these previously discussed \emph{realities} that the Proposal as not sufficiently accounting for is as follows:
\begin{inparaenum}
    \item lack of agreement on what technology like AI really~\textit{is} and what are its capabilities,
    \item  cost and complexity for a business to follow the required regulations,
    \item the known limitations of debiasing techniques and explanations of black boxes,
    \item lack of specifications on how to best implement~\textit{human oversight} in the context of AI systems,
    \item varied and shifting notions of fairness within society,
    \item impact of power imbalances (eg. technological divide, data power, company size, and market share) on the creation and enforcement of and ability to comply with the Proposal.
\end{inparaenum}
\subsection{A Path Forward: Balancing Perspectives}
\subsubsection{\bf{Standardization and Localization}}\label{balance:stand_local}
Standardization may facilitate the translation of fundamental rights, i.e., right to fairness, into standards and specifications to be followed and complied with by all AI actors with the aim of ensuring that AI systems do not discriminate nor mistreat individuals. 

Likewise, localization may allow the clarification of deontological values in more specific and concrete requirements, metrics, or assessments, particular to each enforcement context. This is to prevent a top-down enforcement of operationalizations of fairness that are untenable, or even unfair, in some contexts.
For example, in~\cref{sec:industry} we have summarized the literature demonstrating that ensuring fairness compliance from AI industry could as well be served from a more localized approach to operationalizing fairness. This does not imply the relativization of the legal and ethical principle of fairness but, on the contrary, take into account the wider scenario beyond the purely technical nature of AI and strengthen the enforcement of fairness during the whole life cycle of AI. 

\paragraph{Proposed role of AI Regulation} Standardization should be used to the extent that the measure has a direct link to upholding the deontological value of fairness. In order to ensure the principle of universalization, though, special care must be taken to build in flexible localization allowances. 
\subsubsection{\bf Utilitarian and Egalitarian}
It may be possible to maintain an egalitarian approach to AI Regulations, while also taking advantage of the potential benefits of utilitarian measures.
For example, to promote equality (i.e., bring in egalitarianism) all stakeholders could be given sufficient power to provide inputs on how to maximize and measure their welfare. Any decisions about utilitarian measures would then be based on this input. Note that increased awareness of the use of AI systems and their implications toward fairness among the responding individuals (stakeholders) is essential for a successful process.
This approach would, again, bring up the question of standardization versus localization. Specifically, how highly localized measures would be required to adequately account for the policy expectations of all individuals in an egalitarian fashion. To address this, we would defer to the principles suggested in~\cref{balance:stand_local}. Extensive work is needed to determine how best to implement such a process, but some of the open questions may be best left answered by the inclusive input process itself.

\paragraph{Proposed role of AI Regulation}  The specific framework for how to obtain and incorporate stakeholder inputs should be laid out. A way needs to be found to enforce that \textit{all} stakeholders have sufficient power and influence in AI Regulation decision making processes and that they are themselves sufficiently aware of the potential adverse implications of AI technology.

\subsubsection{\bf Deontological and Consequential}
The EU's stance on fairness is deontological, in that fairness is justified by itself, with no direct subordination to its eventual outcomes. What matters is whether the action is motivated by duty (respect of the~\textit{moral law}: dignity and universalization). However, expectations of individuals on the specifics of what constitutes freedom, equality, and dignity, may vary across cultures, geographies, and contexts. This has led digital and human rights groups to highlight that AI policies should empower individuals, communities, and organisations to contest AI-based systems and to demand redress when they themselves determine that their fundamental rights have been violated \cite{balayanDebiasingRegulatingAI2021}.

The Proposal itself is not intended to legislate individual rights; that is intended to be covered in other laws of the European legal framework. With that in mind, the Proposal could still enforce an individual's need to be informed and to understand the impacts. Therefore transparency, explainability of the design, development and implementaion of AI systems, as well as their output, remains paramount. There must also be understandable and effective methods for stakeholders to adjust the specific standards, such as what uses are forbidden, in the case of unforeseen use cases and impacts or of the recognition of previously ignored violations of the European principles.

\paragraph{Proposed role of AI Regulation}  
Requirements such as documentation and transparency  should specifically serve stakeholders' needs to understand the implications of AI systems for their specific situation, life, and work.
\section{Conclusion}
In this position paper, we presented technical, legal, industrial, sociotechnical, and (moral) philosophical perspectives on the debate on fairness in AI systems with a particular focus on the Proposal of the EU AI Act. We identified a pattern of common arguments representing a key tension in the debate with one side containing \textit{deontological, egalitarian, localized} approaches and the other side containing \textit{standardized, utilitarian, consequential} approaches. We discussed how different (symbolic) ends of the axes could be reconciled and proposed the following roles that the AI Regulation could take to successfully address these tensions:~\textbf{(1)}~apply standardization to uphold deontological values, but ensure universalization by including flexible localization allowances; \textbf{(2)}~lay out a framework to incorporate stakeholder inputs and ensure that they are sufficiently aware of potential adverse implications of AI technology; and \textbf{(3)}~design requirements of documentation and transparency so that they serve the needs of stakeholders.

~\\{\bf Acknowledgements:} This work has received funding from the European Union’s Horizon 2020 research and innovation programme under Marie Sklodowska-Curie Actions (grant agreement number 860630) for the project ``NoBIAS - Artificial Intelligence without Bias'' and (grant agreement number 860621) for the project ``NL4XAI - Natural Language for Explainable AI''.
The authors would like to thank all the panelists of European AI Regulation Week 2021 (\url{https://aia-week.pages.citius.usc.es/}) for their helpful comments on AI regulation.
This work reflects only the authors’ views and the European Research Executive Agency (REA) is not responsible for any use that may be made of the information it contains.

\bibliographystyle{abbrvnat}
\bibliography{refs}

\end{document}